\documentclass[aps,prb,floatfix,twocolumn]{revtex4}
\usepackage{graphicx}
\usepackage{amsmath}
\usepackage{amssymb}
\usepackage{float}
\usepackage{bm}

\newcommand{\ek}{\epsilon_{\bf k}}
\newcommand{\Ek}{E_{\bf k}}
\newcommand{\Deltak}{{\bm \Delta}_{\bf k}}
\newcommand{\xik}{\xi_{\bf k}}

\begin{document}

\title{The Pseudogap Challenge: Understanding the $ab$ Plane~AC Conductivity below $T_c$} 
\author{Andrew Iyengar$^1$, Jelena Stajic$^1$, Ying-Jer Kao$^2$, K. Levin$^1$}
\affiliation{$^1$ James Franck Institute and Department of Physics, University of Chicago, Chicago, Illinois
    60637}
\affiliation{$^2$ Department of Physics, University of Wateroo, Waterloo, ON, N2L3G1, Canada}

\date{\today}
\begin{abstract}

We establish that the cuprate pseudogap plays a crucial role
in the $ab$-plane optical conductivity $\sigma(\omega,T)$
for temperatures $T \le T_c$. 
The pseudogap signatures in $\sigma(\omega,T)$ 
associated
with competing proposals of a ``hidden order''
and ``superconducting'' origin for the $d_{x^2 - y^2}$ pseudogap 
are found to differ in their qualitative $\omega,T$ dependences.
For the latter case, as $T$ increases from $0$, excess low $\omega$ 
weight appears; 
moreover, a much wider range of $\omega$ contributes 
to form the condensate. We discuss these theories in light of
current experiments.
\end{abstract}
\maketitle

While a major fraction of the cuprate phase diagram has been found to be
associated with a pseudogap there is still no definitive explanation
for the origin of this anomalous phase. Here normal state
thermodynamic and transport properties exhibit an excitation
gap of $d$-wave symmetry (with onset at temperature $T^*$)
which appears to be continuous with the
excitation gap of the superconducting phase (with onset at $T_c \le T^*$).
Theoretical schools for addressing this phenomenon fall roughly
into two camps in which
this pseudogap derives from (\textit{i.e.}, is ``intrinsic")
or is independent of (\textit{i.e.,}``extrinsic" to) the superconductivity.
Despite its clear presence in thermodynamical and transport data,
mysteriously, experiments\cite{Timusk} on
the in-plane electrodynamics 
show few conspicuous
signatures of the pseudogap.
In this paper we demonstrate how, in both
theoretical schools, the pseudogap enters as a
rearrangement, relative to BCS predictions,
of spectral weight 
in $\mbox{Re}\:\sigma (\omega) = \sigma_1 (\omega)$.
Moreover, we suggest that it is the width of the range of
frequencies contributing  to form the condensate which provides
a most effective criterion, both for 
descriminating between theoretical approaches, and for
direct comparison with the data.
The experimental observation\cite{vanderMarel,Bontemps2}
that (in the underdoped regime),
relatively high $\omega$ participate in
the condensate poses
a serious challenge 
for both strict BCS theory
and the extrinsic school.

Strict BCS theory
is overly constrained because it contains a single energy scale.
Nevertheless, we argue here that a
mean field theoretic
approach is appropriate because of
(1) the notable similarity of many (but not all)
measured properties to their
BCS predictions.
(2) the very narrow critical regime, and (3) the large separation between
$T^*$ and $T_c$ (which appears to be too big to derive from
fluctuations around strict BCS theory).
Moreover, the electromagnetic response of the superconducting state can
be readily formulated at this generalized mean field level.
In the standard formulation, the imaginary axis conductivity is written as 
$\sigma_{ab}(Q) = \frac{e^2}{i(i\omega_n)}\left[ P_{ab}(Q) - 
n_{ab} \right]$
where $Q= ({\bf q} =0,i\omega_n)$. 
We define Green's functions
\begin{equation}
G(K) = - \frac{i\omega_n^{\prime} + \ek - \mu}{\omega_n^{\prime 2} + \Ek^2} \quad
F(K) = \frac{{\bm \Delta}_{\bf k}}{\omega_n^{\prime 2} + \Ek^2} 
\label{eq:greens}
\end{equation}
\begin{equation}
\Ek^2 = (\ek - \mu)^2 + (\Deltak)^2
\label{eq:dispersion}
\end{equation}
Here $\ek$ refers to a generalized ``band structure'', 
which may differ from the bare hopping dispersion 
$\xik = -2t(\cos k_x + \cos k_y)$, whereas $\Deltak$ 
refers to a generalized ``gap'' parameter, whose ${\bf k}$-dependence 
we separate as $\Deltak = {\bm \Delta} \varphi_{\bf k}$ 
with $\varphi_{\bf k} = \cos(k_x) - \cos(k_y)$ for $d$-wave pairing.

The quasiparticle velocity $v_a = \partial_a \ek$ leads to
\begin{eqnarray}
n_{ab} &=& 
2 \mathop{\sum_{K}} \left[ \partial_a \ek \partial_b \ek (FF - GG)
\right.\nonumber \\  
          &+& \left. 2 \partial_a \ek \partial_b \Deltak FG \right].
\end{eqnarray} 
BCS theory predicts the response 
$P^{\rm BCS}_{ab}(Q) = - 2 \mathop{\sum_{K}} \partial_a \ek \partial_b 
\ek (GG_+ + FF_+)$
{\bf(}where $G=G(K)$, $G_+=G(K+Q)$, and similarly for $F$,{\bf)}
so that the general expression for the superfluid density
$\left(n_s\right)_{ab}  
= n_{ab} - P_{ab}(0)$
reduces in the BCS case (omitting tensor indices) to
\begin{equation}
n_s^{\rm BCS} = 
4 \mathop{\sum_{K}}(
\partial_a \ek \partial_b \ek FF 
+ \partial_a \ek \partial_b \Deltak FG) 
\label{eq:rhos2}
\end{equation}
Since in BCS theory there is only one energy scale, one identifies 
${\bm \Delta}$ as $\Delta_{sc} = \Delta$, and $n_s$ vanishes
only when the excitation gap vanishes. More correctly,
it should be clear that
the superfluid density must necessarily reflect the superconducting
order parameter $\Delta_{sc}$ which may be distinguished from the total
excitation gap $\Delta$, so that $n_s (T) \propto (\Delta_{sc}(T))^2$.
This serves to underline a major shortcoming\cite{Jelena} of strict BCS theory.

The simplest mean field pseudogap approach {\bf(}case (i){\bf)} is referred
to as ``rescaled BCS theory"\cite{Berlinsky}.
Here, the single energy scale framework of BCS theory
$\Delta_{sc} = \Delta$ is preserved 
while modifying the response 
by a frequency-independent factor as
$P_{ab} = \gamma(T) P_{ab}^{\rm BCS}$.
This enhancement causes 
$n_s$ to vanish at $T_c$ lower than $T^*$.
The factor $\gamma(T)$ is related to   
the suppression of the superfluid $n_s = f(T)n_s^{\rm BCS}$ 
by the sum rule constraint
$\gamma(T) = \frac{n}{n-n_s^{\rm BCS}(T)}{\bf(}1 - f(T){\bf)} + f(T)$,
with $f(T)$ chosen to vanish at $T_c$ and fitted to
$n_s(T)$.
The microscopic physics of this otherwise
phenomenological scheme 
is presumably connected to the approach
of Lee and Wen\cite{LeeWen}; $\gamma$ may also reflect the renormalization 
of electron charge as can occur in Fermi-liquid based approaches to the 
superconducting state\cite{Millis_review}.
In overdoped samples with $T_c = T^*$, BCS theory works well, i.e.,
$f(T) = 1$.

Alternatively, one may argue that $\Delta_{sc} \ne \Delta$, so 
that at the mean field level, the gap and the order parameter are distinct. 
The extrinsic school introduces a pseudogap $\Delta_{pg}$ directly into
the quasi-particle bandstructure $\ek$ and defines
${\bm \Delta} \equiv \Delta_{sc}$ in the above mean field formalism, yielding 
$n_s \propto \Delta_{sc}^2$ via Eq.~(\ref{eq:rhos2}). Below $T_c$, 
Eq.~(\ref{eq:dispersion}) shows that in this school 
the dispersion of the fermionic quasi-particles 
has no simply defined BCS-like excitation gap. 
However, in this paper (as elsewhere\cite{Loram}) we will
assume that one can define a meaningful excitation gap
for the extrinsic case as 
$\Delta_{\bf k } \approx \sqrt {(\Delta_{\bf k }^{pg}) ^2 + 
(\Delta_{\bf k } ^{sc} )^2}$.
We implement this picture 
by introducing
an additional order parameter which breaks translation symmetry, 
leading to two bands
$\epsilon_{\bf k}^{\pm} = \pm \sqrt{\xik^2 + (\Delta^{pg}_{\bf k})^2}$.
The precise superfluid density
expression is slightly more complicated than 
Eq. ~(\ref{eq:rhos2}) due to interband effects.\cite{DDW_rhos}
While we calculate $P_{ab}$ for a $d$-density wave\cite{Laughlin,Nayak}
(DDW) here, 
we argue that our results are
generic, applying to other extrinsic pseudogap 
theories.\cite{Loram,Nozieres2}
The Green's functions $F^{\pm},G^{\pm}$ for the two 
bands are obtained from
Eq.~(\ref{eq:greens}),~(\ref{eq:dispersion}) by replacing $\epsilon _{\bf k}$
with $\epsilon ^{\pm} _{\bf k}$. 
For the $d$-density-wave/superconducting coexistent state, we find
\begin{eqnarray}
P_{ab}(Q) &=&  - 2 \mathop{\sum'_{K,\nu=\pm}} \left[
v^{(1)}_a v^{(1)}_b (G^{\nu}G_+^{\nu} + F^{\nu}F_+^{\nu}) \right. \\
\nonumber
&+& \left. v^{(2)}_a v^{(2)}_b (G^{\nu}G_+^{-\nu} + F^{\nu}F_+^{-\nu}) \right]
\label{eq:extrinsic}
\end{eqnarray}
where momentum is summed over the half Brillouin zone, 
$\nu$ is the band index, 
$v^{(1)}_a = \partial_a \epsilon_{\bf k}$, and 
$v^{(2)}_a = (\xi_{\bf k} \partial_a \Delta^{pg}_{\bf k} 
- \Delta^{pg}_{\bf k} \partial_a \xi_{\bf k})/\epsilon_{\bf k}$.

Finally, we consider a third mean field alternative (iii)
in which the pseudogap is \textit{intrinsically} connected with
the superconductivity\cite{Chen2}. 
In this picture $\Delta_{pg}$ is associated with a stronger-than-BCS
attractive interaction, so that (finite momentum)
pairs form at a higher temperature
$T^*$ than that at which they Bose condense ($T_c$). Here the pseudogap
enters on the same footing as the superconducting order parameter,
and the fermionic dispersion is given by Eq.~(\ref{eq:dispersion}) with
the value of $\Deltak$ assigned to
be $\Delta_{\bf k}$, and $\ek = \xik$. 
In this last approach, which has a detailed 
microscopic basis\cite{Chen2},
the superfluid density has the form
\begin{equation}
n_s = [\Delta_{sc}^2/\Delta^2] n_s^{\rm BCS}; ~~\Delta = \sqrt {(\Delta^{pg}) ^2 + 
(\Delta^{sc} )^2}.
\label{eq:rhos4}
\end{equation}  
The corresponding electromagnetic response is
the sum of fermionic (BCS, with full gap $\Delta(T)$)
and bosonic contributions 
\begin{eqnarray}
\nonumber P_{ab}(Q)&=& - 2 \mathop{\sum_K}
\partial_a\xik\partial_b\xik(FF_+ + GG_+) \\
&+& A \mathop{\sum_P} \partial_a\Omega_{p}\partial_b\Omega_{p} 
t^{pg}(P) t^{pg}(P+Q)
\label{eq:intrinsic}
\end{eqnarray}
Here we have approximated\cite{Varlamov} the second bosonic term, which 
originates from an Aslamzov-Larkin diagrams)\cite{AL_diagrams}.
It represents
the coupling of radiation to uncondensed pairs of charge-2$e$
responsible for suppressing $n_s$ relative to $n_s^{BCS}$. The pairs have a
propagator
$t_{pg}(P)$, which at small $P$ can be
parameterized as 
$t_{pg}^ {-1}(p,\omega) = \omega(1+ i \nu) - \Omega_p$ (after 
continuation to real frequency $\omega$.) Here
$\Omega_p = p^2/2M$ and the pair chemical
potential is zero for $ T \le T_c$.  While our microsopic theory\cite{Kosztin1}
yields the pair mass $M$ and lifetime parameter associated with $\nu$, 
the diagrammatic calculations can be avoided by borrowing
from general TDGL formulations\cite{Schmidt} which describe the hydrodynamics
of non-condensed bosons. In particular,
the optical sum rule constrains the conductivity spectral weight 
of this boson contribution to be 
$n_s^{\mbox{boson}} = [\Delta_{pg}^2/\Delta^2] n_s^{\rm BCS}$, 
determining the prefactor $A$.

In the intrinsic school 
the condensate ($\Delta_{sc}$) grows with decreasing $T$ at the expense
of $\Delta_{pg}$ as the
fraction of (quasi-ideal Bose gas) finite momentum pair excitations decreases.
Precisely at $T=0$, $\Delta_{pg}$ vanishes and the ground state
is given by the BCS generalization
first proposed in Ref. \onlinecite{Leggett}.
By contrast,
in the extrinsic model $\Delta_{pg}$ is relatively $T$ independent
below $T_c$ in the underdoped regime.
One can summarize: to a good approximation,
for the intrinsic model\cite{Chen2} {\bf(}case (iii){\bf)}
$\Delta_{pg}^2 (T) \approx \Delta^2 (T_c) (T/T_c)^{3/2}$,
whereas for the extrinsic model {\bf(}case (ii){\bf)} at low doping
$\Delta_{pg}^2 (T) \approx \Delta_{pg}^2 (T_c)$.  

The following calculations for $\sigma_1 (\omega)$ address
an underdoped (UD) 
and a slightly over-doped (OD) cuprate
with hole concentration $x$
quantified by excitation gap ratio $\Delta(T_c)/\Delta(0)$ which
assumes the values
0.99 and 0.5 respectively. 
This corresponds roughly to $x \approx 0.06$
and $ x \approx .17$.  The two quantities
$n_s(0) /\Delta(0)$ and $T_c/T^*$ were computed within each microscopic
theory and roughly fit to experiment. We presume that the temperature
dependence of the full excitation gap for $T \le T_c$ is given by the
BCS $d$-wave result, so that $\Delta(0) / T^*$ is a universal
number, as seems to be the case experimentally\cite{Oda}.
We include scattering from random non-magnetic impurities by 
renormalizing
$F$ and $G$ in the above formulae,\cite{Ykao_AG} and, for simplicity,
compute the bosonic term in Eq(7) using standard TDGL results\cite{Schmidt}.
Following conventional procedures, 
\cite{Fehrenbacher} we pre-calculate a table of self-energies 
on the real frequency axis. 
We handle the complicated behavior of energy and ${\bf k}$ integrands
using automatic 1d or 2d adaptive quadrature, 
determining $\sigma _1$ to 1\% relative accuracy.
While we have explored both Born and unitary limits, the computations 
are more tractible in the unitary limit, which we consider here with
scattering potential $u\:\delta({\bf r})$, $u/2t \rightarrow \infty$,
and with 0.01 impurities per unit cell.
Our numerical
calculations (1) conserve the integrated density of
states, (2) conserve conductivity spectral weight, 
and (3) yield 
\cite{Lee_limit}  $\sigma_1 (T=0,\omega=0) = (2/\pi) 2t/{\bm \Delta}$.
Here and throughout we use conductance units $e^2/h=1$. 

\begin{figure}[!thb] 
\centerline{\includegraphics[angle=0,width=3.4in
]{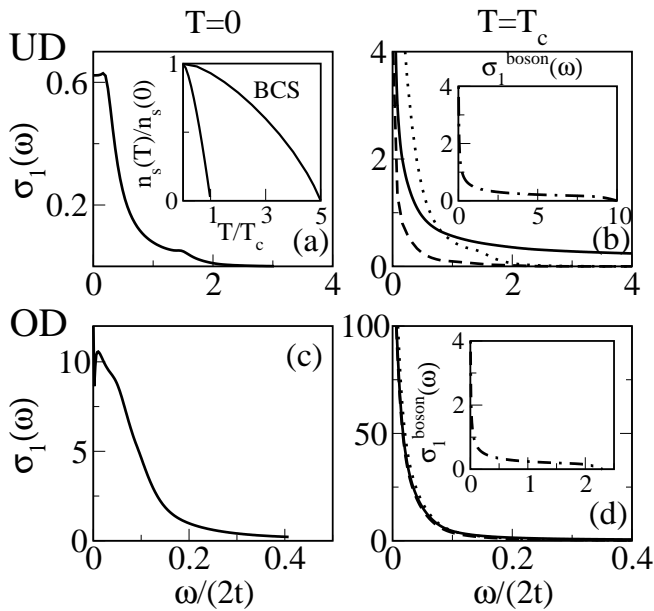}}   
\caption{Real part of ac conductivity for the intrinsic (solid),
rescaled BCS (dotted) and strict BCS (dashed) models
at underdoped (UD) and slightly
overdoped (OD) $x$  at $T =0$ (1a, 1c) and $T=T_c$ (1b,1d). Insets
plot the corresponding $n_s$ compared to $n_s^{\rm BCS}$ (1a) and
the bosonic contribution to $\sigma_1$ (1b).}
\label{fig:intrinsic}
\end{figure} 
\begin{figure}[!thb]
\centerline{\includegraphics[angle=0,width=3.4in
]{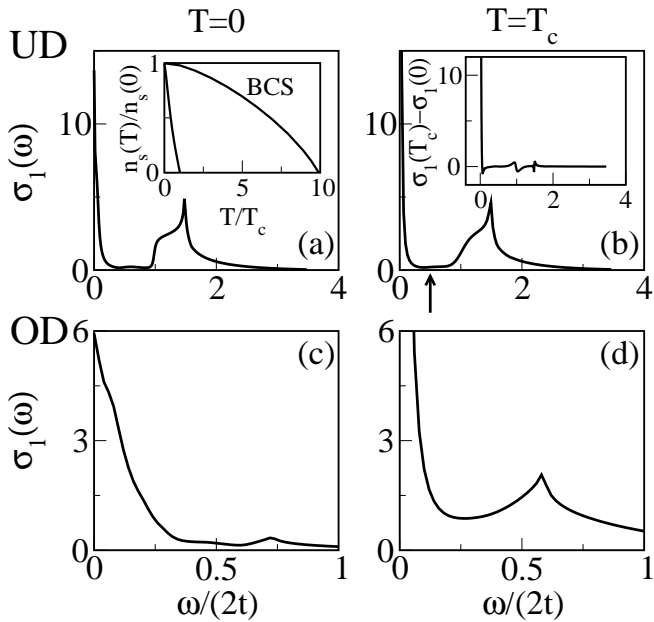}}
\caption {$\sigma_1(\omega)$ for the extrinsic psudogap
at underdoped (UD) and slightly overdoped (OD) $x$ at $T=0$
(2a,2c) and $T=T_c$ (2b,2d).  The arrow indicates 
$\Delta(0) \approx \Delta_{pg}(T_c)$}.
\label{fig:extrinsic}
\end{figure}
\begin{figure}[!thb]
\centerline{\includegraphics[angle=0,width=3.2in
]{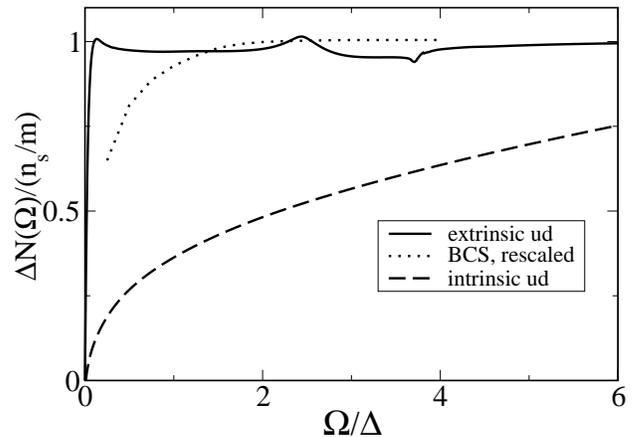}}
\caption{Partial integral of condensing spectral weight, normalized by the 
total, for the three pseudogap models (underdoped).}  
\label{fig:rhos}
\end{figure}
In the insets of Figures 1a and 2a are plotted 
the normalized superfluid densities for the
intrinsic (iii) and extrinsic cases (ii) in the underdoped regime.
The  strict ``BCS" curves 
for each
case are chosen to have
the same excitation gap parameter, so that
$n_s^{\rm BCS}$ vanishes at $T^*$. 
It is clear from the above discussion
that the calculated superfluid density
is depressed and vanishes at a lower temperature ($T_c$) than its
BCS counterpart. Sum rule considerations thus indicate that
\textit{there must be excess spectral weight in $\sigma_1 (\omega , T)$
at $\omega \ne 0$ associated with pseudogap effects}. 

The frequency distribution of this
excess spectral weight can 
be inferred from 
the plots in Figures 1  and 2 which show
$\sigma_1 (\omega)$
for the UD and OD hole concentrations
at $T=0$ and $T=T_c$ in the intrinsic (solid lines)
and extrinsic cases respectively. 
In Figures 1a, and 1c the rescaled BCS model (dotted lines) 
coincides with the solid curves at $T=0$
due to the complete condensation 
of the pseudogap.
It is clear that in the intrinsic case the added pseudogap weight
(present at any $T \ne 0$) appears at very low $\omega$, varying
as\cite{Schmidt} $1/\sqrt{\omega}$, as well as at very high $\omega$
where it introduces a long tail associated with the fermion induced
damping of non-condensed pairs.  Interactions with the
fermionic two particle continuum (which itself is gapped)
constrains the bosonic lifetime. By contrast, in
the extrinsic case the pseudogap is associated with
a secondary maximum 
which appears at the energy scale 
$\Delta_{pg} (T_c) \approx \Delta(0)$. 
This
arises from interband effects
\cite{Nayak} with threshold at $\omega = 2 \mu$.

It should be noted that a direct comparison of $\sigma_1(\omega)$
between theory
and experiment is complicated because the origin of the so-called
mid-infrared spectral weight is uncertain\cite{Tanner}, and because the ratio
of the $T=0$ condensate to the total integrated spectral
weight (at $T_c$) is anomalously small (20\% or less)\cite{Tanner}.
Thus, 
we focus on the \textit{difference}
curves, and their associated
integrals. 
We introduce 
the partially integrated spectral weight
$\Delta N(\Omega)=\frac{1}{\pi}\int _0 ^{\Omega}
[ \sigma_1(\omega,T_c) -\sigma_1(\omega, 0)] d\omega$, which 
is plotted in Figure 3 for the underdoped regime.
This plot indicates which frequencies are required
to account for the superfluid density in both the extrinsic and
intrinsic cases. 

Strikingly, the intrinsic
pseudogap and rescaled BCS theories both show substantial contributions to the 
condensate up to relatively high frequencies of order 4 $\Delta(0) 
\approx 960 $cm$ ^{-1}$ or more.
By contrast,
in the extrinsic school, the condensate derives from
a much narrower range of frequencies. Since
interband conductivity terms make a negligible condensate contribution, this
observation holds even for extrinsic theories which lack
interband 
effects\cite{Loram,Nozieres2}.
An analogous narrowing of condensate frequencies occurs in strict BCS
theory at low $x$ when $\Delta(T)$ is properly chosen to
be $T$ independent, so that
the only relevant energy scales are
the much lower $T_c$ and scattering rate.

For $T \le T_c$,  inelastic scattering, which has
been ignored in the above calculations, may affect any
direct comparison between theory and experiment. This leads to
non-monotonic $\omega$ dependences and is expected
to be most important in
the optimal and over-doped regime, where inelastic effects
are strongly $T$ dependent below $T_c$, reflecting the $T$ dependence
of $\Delta(T)$.
It follows, similarly, that 
in the underdoped regime
where $\Delta(T)$ is constant, inelastic effects
should not contribute appreciably to $\Delta N $.

Finally, it can be noted that the only, 
nevertheless ambiguous\cite{Bontemps}, evidence for a pseudogap appears in
the experimental literature\cite{Timusk}
via Drude fits to $\sigma(\omega)$,
where the resulting $1/\tau (\omega)$ is suppressed at 
frequency $\omega_o \approx 600$ cm$^{-1}$. Because of its $x$
independence we speculate that $\omega_o$ may reflect a phonon
or other boson\cite{Timusk}
with coupling modulated by the opening of an excitation gap. Although
these secondary pseudogap
effects are beyond the scope of this paper, we find
here that Drude fits are quite
generally problematic for cases (ii) and (iii) discussed
here. 

In summary, in this paper (i) we have investigated competing
proposals for the pseudogap origin and have presented theoretical
(and rather generic) mean field expressions for $\sigma(\omega)$.
(ii) Both major mean field pseudogap schools 
lead to the prediction
that pseudogap effects must enter as a rearrangement of
\textit{excess} spectral weight (relative to strict BCS theory predictions)
in the ac conductivity. For the intrinsic school, there is 
a rapid growth in $\sigma_1$ with increasing low $T$, at fixed
low frequencies (which may have been observed experimentally\cite{Bonn}), 
as well as long high $\omega$
tails, (possibly also observed\cite{vanderMarel,Bontemps2})
which tails become more pronounced with
increasing $T$.  While for the extrinsic school, there is
a transfer of spectral weight from low to high $\omega$. However,
here the resulting
secondary maximum is present at all $T \le T^*$ and does not
participate in the condensation. Direct comparison with experimental
data is made complex by the small integrated spectral weight
in the condensate and our neglect of inelastic scattering\cite{Tanner}. 
Therefore, we have
focused on the condensing portion of the spectral weight 
by considering the \textit{change} in 
$\sigma_1(\omega)$ from $0$ to $T_c$.
(iv) We find that when $\Delta(T) \approx$ const, strict
BCS theory cannot account for the relatively high ($\sim 4\Delta(0)$)
frequency
scale associated with the condensate. The
extrinsic approach suffers from a similar problem.
In an artificial way rescaled BCS theory remedies this weakness.
However, the observed
high frequency contributions arise naturally 
in an intrinsic model; their presence reflects
the underlying scattering of
non-condensed bosons from the two particle fermionic states.

This work was supported by NSF-MRSEC Grant No. DMR-9808595 (AI, JS, KL)
and by NSERC of Canada and Research Corporation (YK).
We especially thank Qijin Chen for useful discussions and
are grateful to D. Basov, N. Bontemps, J. Carbotte and Shi-Na Tan
for multiple conversations, and to
W. Putikka and E. Schachinger for consultation regarding numerical
techniques.

\bibliographystyle{prsty}
\bibliography{/home/apiyenga/Wrups/ACconductivity/Paper/sigma}

\end{document}